\documentclass[twocolumn,floatfix,preprintnumbers,nofootinbib,superscriptaddress]{revtex4-1}

\usepackage{ulem}
\usepackage{bm}
\usepackage{times}
\usepackage{amssymb,amsbsy,amsmath,amsfonts}
\usepackage{graphicx}
\usepackage{float}
\usepackage{color}
\usepackage{morefloats}
\usepackage{rotating}
\usepackage{srcltx}
\usepackage{slashed}
\usepackage{subfigure}
\usepackage{multirow}
\usepackage{verbatim}
\usepackage{hyperref}
\usepackage{tabularx}
\usepackage{braket}
\usepackage{diagbox}
\usepackage{appendix}
\usepackage{subfigure}
\usepackage{makecell}
\usepackage{booktabs}
\usepackage[section]{placeins}

\usepackage[outdir=./]{epstopdf}
\allowdisplaybreaks
\begin{document}
\title{Strong decays of $a_0$, $f_0$, $f_2$, and $K^*_2$ resonances as dynamically generated states of two vector mesons}

\date{\today}

\author{Qing-Hua Shen}~\email{shenqinghua@impcas.ac.cn}
 \affiliation{Institute of Modern Physics, Chinese Academy of Sciences, Lanzhou 730000, China}
 \affiliation{School of Physical Science and Technology, Lanzhou University, Lanzhou 730000, China}
 \affiliation{School of Nuclear Sciences and Technology, University of Chinese Academy of Sciences, Beijing 101408, China}
 
\author{Li-Sheng Geng} \email{lisheng.geng@buaa.edu.cn}
\affiliation{School of Physics, Beihang University, Beijing 102206, China}
\affiliation{Peng Huanwu Collaborative Center for Research and Education, Beihang University, Beijing 100191, China}
\affiliation{Beijing Key Laboratory of Advanced Nuclear Materials and Physics, Beihang University, Beijing, 102206, China}
\affiliation{Southern Center for Nuclear-Science Theory (SCNT), Institute of Modern Physics, Chinese Academy of Sciences, Huizhou 516000, China}

 \author{Ju-Jun Xie}~\email{xiejujun@impcas.ac.cn}
\affiliation{Institute of Modern Physics, Chinese Academy of Sciences, Lanzhou 730000, China}
\affiliation{School of Nuclear Sciences and Technology, University of Chinese Academy of Sciences, Beijing 101408, China}
\affiliation{Southern Center for Nuclear-Science Theory (SCNT), Institute of Modern Physics, Chinese Academy of Sciences, Huizhou 516000, China}
 
 \begin{abstract}
 
The two-body strong decays of the $f_0(1500)$, $f_0(1710)$, $a_0(1710)$, $f_2(1270)$, $f_2'(1525)$, and $K_2^*(1430)$ resonances are investigated, assuming them as dynamically generated states of two vector mesons via $s$-wave interactions. The partial decay widths of all the possible two-body pseudoscalar meson-pseudoscalar meson final states are calculated considering the triangular diagrams. It is found that the ratios of branching fractions are similar to the previous results for most channels, which were obtained by using the real-axis method and considering the box diagrams. However, in this work, our focus is on the partial decay widths and their ratios. More precise experimental measurements are needed to test the model calculations and determine the nature of these scalar and tensor mesons. It is anticipated that the BES\uppercase\expandafter{\romannumeral3}, Belle\uppercase\expandafter{\romannumeral2} and LHCb collaborations will conduct these measurements in the future.

\end{abstract}

\maketitle

\section{Introduction}

The study of hadrons, particularly exotic hadrons, is one primary research field of particle physics. In the classical quark models, mesons are composed of pairs of quark and antiquark ($q\bar{q}$) and baryons are made up of three quarks ($qqq$). However, from 2003, some exotic hadrons were discovered experimentally and many theoretical investigations contributed to them [see reviews in Refs.~\cite{Liu:2013waa,Hosaka:2016pey,Chen:2016qju,Lebed:2016hpi,Guo:2017jvc,Olsen:2017bmm,Brambilla:2019esw}]. Following these discoveries, many models have been proposed to explain the nature of these exotic hadrons, e.g., hadronic molecules, multiquark states or hybrid states. The hadronic molecular picture has been widely used to interpret most newly discovered hadronic states, such as the $XYZ$ particles~\cite{Yang:2021sue,Wang:2022xga,Liu:2020nil,Liu:2019zvb,Wang:2018djr,Wang:2017mrt,Aceti:2014uea} and the $P_c$ pentaquark states~\cite{Wang:2023iox,Wang:2022mxy,Wang:2021crr,Shen:2016tzq,Lu:2015fva}.

In the light-quark sector, the scalar mesons $f_0(1500)$, $f_0(1710)$, and  $a_0(1710)$, and the tensor mesons $f_2(1270)$, $f_2'(1525)$, and $K_2^*(1430)$ have attracted a lot of attention. The $f_0(1370)/f_0(1500)$ and $f_0(1710)$ states were previously considered as glueball candidates~\cite{Giacosa:2005zt,Chanowitz:2005du,Chao:2007sk,Li:2021gsx}. Using the chiral unitary approach~\cite{Oller:1997ti,Oset:1997it,Chiang:1997di,Oller:1998hw,Inoue:2001ip,Jido:2003cb,Roca:2005nm,Geng:2006yb,Gamermann:2006nm,Gamermann:2007fi,Molina:2008jw,Geng:2008gx,Sarkar:2010saz,Xiao:2013jla,Liang:2014kra,Zhou:2014ila,Lu:2014ina,Yu:2019yfr,Molina:2009eb,Molina:2009ct,Oset:2010tof,Molina:2010tx,Xiao:2013yca}, the vector-vector interaction has been investigated in a coupled-channel approach, dynamically generating the $f_0(1370)$ and $f_2(1270)$ mesons~\cite{Molina:2008jw,Geng:2008gx,Geng:2016pmf,Molina:2019rai}, which are naturally explained as bound states of $\rho \rho$. Similarly, the $f_0(1710)$ and $f_2'(1525)$ state can also be obtained and are primarily coupled to the $K^*\bar{K}^*$ channel in the spin $J=0$ and $J=2$ sectors, respectively. Additionally, the $K_2^*(1430)$ state also appears~\cite{Geng:2008gx}. The molecular nature of $f_0(1370)$, $f_0(1710)$, $f_2(1270)$, $f_2'(1525)$, and $K_2^*(1430)$ has been successfully tested in numerous processes~\cite{Nagahiro:2008um,Branz:2009cv,MartinezTorres:2009uk,Geng:2010kma,Dai:2013uua,Dai:2015cwa,Xie:2014gla,Xie:2014twa}. The $a_0(1710)$ state~\footnote{In the molecular picture, its pole mass is about 1780 MeV. In this work, we denote it as $a_0(1710)$.} can be viewed as the isospin-one partner of the $f_0(1710)$~\cite{Geng:2008gx,Du:2018gyn}. Similar conclusions have been obtained in Refs.~\cite{Garcia-Recio:2010enl,Garcia-Recio:2013uva,Wang:2019niy,Wang:2021jub,Wang:2022pin}, where the pseudoscalar-pseudoscalar coupled channels were considered. The  $a_0(1710)$ state productions in the $D^+_s \to \pi^+ K^0_S K^0_S$ and $D^+_s \to \pi^0 K^0_S K^+$ decays were studied in Refs.~\cite{Dai:2021owu,Zhu:2022wzk,Zhu:2022guw} in the molecular picture where it couples mostly to the $K^*\bar{K}^*$ channel. On the other hand, in the molecular picture, the two-photon productions of $a_0(1405)$, $a_0(1710)$, $f_0(1500)$, and $f_0(1710)$ resonances were calculated within the effective Lagrangian model in Ref.~\cite{Yang:2024wzp}, where it is pointed out that confirming $a_0(1450)$ and $a_0(1710)$ as the isovector partners of $f_0(1500)$ and $f_0(1710)$ is crucial in determining their nature. We note that, Ref.~\cite{Guo:2022xqu} argued that the $a_0(1710)$ (denoted as $a_0(1817)$ in that reference) could be a good isovector partner of the $X(1812)$ state, and the possibility of the $f_0(1710)$ as the scalar glueball cannot be excluded.

Using the real axis method and considering contributions from the box diagrams, the two-body strong decay ratios of these vector meson-vector meson molecular states, i.e., $f_0(1370)$, $f_0(1710)$, $f_2(1270)$, $f_2'(1525)$, and $K_2^*(1430)$, were calculated~\cite{Geng:2009gb}. With the resonance chiral theory approach, the masses and decay widths of the lowest tensor nonet, including $f_2(1270)$, $f_2'(1525)$, and $K_2^*(1430)$, were calculated in Ref.~\cite{Chen:2023ybr} by incorporating higher-order resonance chiral operators. Along this line, in this work, we revisit the two-body strong decays of $R \to PP$, with $P$ representing the light-flavor pseudoscalar mesons and $R$ representing the scalar and tensor mesons $f_0(1370)$, $f_0(1710)$, $a_0(1710)$, $f_2(1270)$, $f_2'(1525)$, and $K_2^*(1430)$. This involves considering triangle diagrams, where they first couple to a vector-vector pair, which then decay into two pseudoscalar mesons by exchanging a pseudoscalar meson with two vector-pseudoscalar-pseudoscalar ($VPP$) vertices. We would like to mention that the method used here to calculate the partial decay widths of $R \to PP$ is equivalent to the one used in Refs.~\cite{Molina:2008jw,Geng:2008gx} with the box diagrams which were added to the potential, and it was found that the contribution to the real part of the potential of the box diagram is very small.

This article is organized as follows. In the next section, we present the relevant theoretical formalism for the vector meson-vector meson interactions and the reaction mechanisms for the $R \to VV$ decays. In Sec.~\ref{numericalresults}, the numerical results for the partial decay widths and their ratios are shown. Discussions and comparisons with previous calculations and available experimental data are also presented. Finally, a short summary is given in Sec.~\ref{Summary}.

\section{Theoretical Formalism}~\label{formalism}

The relevant ingredients in calculating the partial decay widths of the $a_0$, $f_0$, $f_2$ and $K^*_2$ resonances as dynamically generated states of two vector mesons to a pair of pseudoscalar mesons are the strong couplings of $R$ to the $VV$ channel, which can be extracted from the two-body vector-vector unitarized $s$-wave scattering amplitudes of the chiral unitary approach. Although the forms of these interactions have been detailed in Refs.~\cite{Molina:2008jw,Geng:2008gx}, we briefly review them in this section. This will allow us to review the general procedure of obtaining the two-body couplings entering the partial decay width calculations. 

\begin{figure}[htbp]
   \centering
   \includegraphics[scale=1.2]{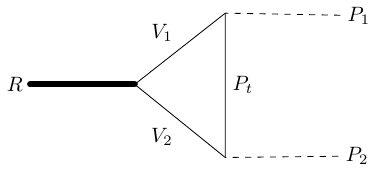}
   \caption{Diagrammatic representation of two pseudoscalar mesons decay mechanism for scalar or tensor resonances dynamically generated by the interactions of two vector mesons in coupled channels.}
   \label{fig:triangular}
\end{figure}

The scalar or tensor resonances ($R$) that are dynamically generated from the vector meson-vector meson interaction in coupled channels have strong couplings to the vector-vector channels, thus they decay mainly into two pseudoscalar mesons with the triangular mechanism as shown in Fig.~\ref{fig:triangular}, where they firstly couple to two vector mesons $V_1$ and $V_2$, then the $V_1$ and $V_2$ transform into two pseudoscalar mesons $P_1$ and $P_2$ by exchanging one pesudoscalar meson $P_t$. Following Refs.~\cite{Branz:2009cv,Yamagata-Sekihara:2010mia,Molina:2010bk,Xie:2014twa,Xie:2015isa}, the effective interaction vertex of $RVV$ can be written as
\begin{equation}
t_{RV_1V_2}=g_{RV_1V_2}\mathcal{P}_{J}(V_1V_2),
   \label{eq:RVV}
\end{equation}
where $\mathcal{P}_J$ is the spin projection operator, which projects the two vector mesons $V_1$ and $V_2$ into spin $J$. The $\mathcal{P}_0$ and $\mathcal{P}_2$ can be written as~\cite{Molina:2008jw,Geng:2008gx}
\begin{eqnarray}
\mathcal{P}_0(V_1V_2) &=& \frac{1}{\sqrt{3}}\epsilon_i(V_1) \epsilon_i(V_2), \\
    \mathcal{P}_2(V_1V_2) &=& \frac{1}{2} \left[ \epsilon_i(V_1) \epsilon_j(V_2)+\epsilon_j(V_1) \epsilon_i(V_2) \right ] \nonumber \\
    && -\frac{1}{3}\epsilon_m(V_1) \epsilon_m(V_2)\delta_{ij},
\end{eqnarray}
where $\epsilon_i(V_{1,2})$ is the polarization vector of the vector mesons $V_{1,2}$. For a vector meson with mass $m$ and three-momentum $\vec{k}$, they can be written in a compact form as~\cite{Greiner:1996zu}

\begin{equation}
   \begin{aligned}
      \epsilon(\vec{k},\lambda)=\left( \frac{\vec{k} \cdot \epsilon_{\lambda} }{m}, \epsilon_{\lambda}+\frac{\vec{k} \cdot \epsilon_{\lambda}}{m(k^0+m)}\vec{k} \right),
   \end{aligned}
\end{equation}
where $k^0 = \sqrt{|\vec{k}|^2 + m^2}$ and $\epsilon_{\lambda}$ with $\lambda = 0$ and $\pm 1$ are taken as 
\begin{equation}
   \begin{aligned}
      \epsilon_0=
      \begin{pmatrix}
       0\\
       0\\
       1  
      \end{pmatrix}, \hspace{0.8cm}
      \epsilon_{\pm 1}=\frac{\mp 1}{\sqrt{2}}
      \begin{pmatrix}
         1\\
         \pm i\\
         0
      \end{pmatrix}.
   \end{aligned}
\end{equation}
\noindent

In Eq.~(\ref{eq:RVV}), $g_{RV_1V_2}$ is the coupling constant for the $RV_1V_2$ vertex. One can extract the effective strong coupling of the resonance $R$ to the $VV$ channel at the pole position of $R$ from the unitarized scattering amplitude $T_{VV \to VV}$, which was obtained by solving the Bethe-Salpeter equation and considering the re-scattering of all the coupled channels. In Ref.~\cite{Geng:2008gx}, the dimensional regularization method with $\mu = 1000$ MeV is used to regularize the vector-vector loop functions in the chiral unitary approach. By reproducing the $f_2(1270)$, $f_2'(1525)$, and $K_2^*(1430)$ resonances, the subtraction constant $a_\mu$ is determined for different isospin channels. We show them in Table~\ref{tab:ac}. More details can be found in Ref.~\cite{Geng:2008gx}.

\begin{table}[htbp]
    \centering
    \renewcommand\arraystretch{1.5}
   \tabcolsep=3pt
    \caption{Subtraction constants $a_\mu$ for different vector-vector channels with $\mu = 1000$ MeV.}
    \begin{tabular}{c|c|c}
      \toprule[1pt]
      \toprule[1pt]
      sector& channel & $a_\mu$\\
      \hline
      \multirow{3}{*}{strangeness=0}&$\rho \rho$ & $-1.636$ \\
      \cline{2-3}
      &$K^*\bar{K}^*$ & $-1.726$ \\
      \cline{2-3}
      &$\omega \omega$, ~$\omega \phi$, ~$\phi \phi$, ~$\rho \omega$, ~$\rho \phi$ & $-1.65$ \\
      \hline
      strangeness=1& $\rho K^*$, ~$K^* \omega$, ~$K^* \phi$ & $-1.85$\\
      \hline
      strangeness=2& $K^* K^*$ & $-1.726$\\
      \bottomrule[1pt]
      \bottomrule[1pt]
    \end{tabular}
    
    \label{tab:ac}
\end{table}

Most of these coupling constants for $f_2(1270)$, $f_2'(1525)$, $f_0(1710)$, and $K_2^*(1430)$ resonances in the isospin basis were already obtained in Ref.~\cite{Geng:2008gx}, which are complex. Following the same procedure as in Ref.~\cite{Geng:2008gx} but with the cutoff method matching with the regularization method used in Ref.~\cite{Geng:2008gx}, we obtain the pole positions and the couplings for $a_0(1710)$ and $f_0(1500)$. All the obtained pole positions and coupling constants for $RVV$ vertices are listed in Table~\ref{Tab:coupling}. Note that the coupling constants for $a_0(1710)$ and $f_0(1500)$ resonances are determined at their pole positions $(1780,-66i)$ MeV and $(1512,-26i)$ MeV, respectively. 

\begin{table*}[htbp]\footnotesize
   \centering
   \renewcommand\arraystretch{2}
   \tabcolsep=4pt
   \caption{The obtained pole positions and couplings for the vector meson-vector meson dynamically generated states in the isospin basis. All quantities are in units of MeV.}
   \begin{tabular}{ccccccc}
      \toprule[1pt]
      \toprule[1pt]
      \multicolumn{7}{c}{strangeness=0 and isospin=0}\\
      \hline
     & pole position & $K^*\bar{K}^*$ & $\rho \rho$ & $\omega \omega$ & $\omega \phi$ & $\phi \phi$ \\
      \hline
      $f_2(1270)$ & $(1275,-i)$&$(4733,-53i)$ & $(10889,-99i)$ & $(-440,7i)$ & $(777,-13i)$ &$(-675,11i)$\\
      $f_2'(1525)$ &$(1525,-3i)$& $(10121,101i)$ & $(-2443,649i)$ & $(-2709,8i)$ & $(5016,-17i)$ & $(-4615,17i)$\\
      $f_0(1710)$ & $(1726,-14i)$&$(7124,96i)$ & $(-1030,1086i)$ & $(-1763,108i)$ & $(3010,-210i)$ & $(-2493,-204i)$\\
      $f_0(1500)$& $(1512,-26i)$&$(1208,-419i)$ & $(7906,-1084i)$ & $(-40,i30)$ & $(34,-42i)$ & $(12,24i)$\\ 
      \hline
      \multicolumn{7}{c}{strangeness=0 and isospin=1}\\
      \hline
      &pole position & $K^* \bar{K}^*$ & $\rho \omega$ & $\rho \phi$ \\
      \hline
      $a_0(1710)$&$(1780,-66i)$ & $(7526,-1525i)$  & $(-4042,1389i)$ & $(4998,-1869i)$\\
      \hline
      \multicolumn{7}{c}{strangeness=1 and isospin=1/2}\\
      \hline
      &pole position & $\rho K^*$ & $K^* \omega$ & $K^*\phi$ & & \\
      \hline
      $K_2^*(1430)$&$(1431,-i)$ & $(10901,-71i)$ & $(2267,-13i)$ & $(-2898,17i)$ & & \\
      \bottomrule[1pt]
      \bottomrule[1pt]
   \end{tabular}
   \label{Tab:coupling}
\end{table*}

There are two scalar $f_0$ states, the higher one, $f_0(1710)$, mostly couples to the $\bar{K}^* K^*$ channel, and the lower one is mostly a $\rho \rho$ bound state. Previously, the lower one was considered as the scalar meson $f_0(1370)$, with estimated mass and width $1200 \sim 1500$ MeV and $200 \sim 500$ MeV, respectively, as quoted in the Review of Particle Physics (RPP)~\cite{ParticleDataGroup:2022pth}. The $f_0(1500)$ has a mass of $1522 \pm 25$ MeV and a width of $108 \pm 33$ MeV. In Ref.~\cite{Wang:2019niy}, the $\rho \rho$ scattering was calculated considering the coupled channels of pseudoscalar mesons. It was found that the dynamically generated state $f_0$ is more consistent with $f_0(1500)$ rather than $f_0(1370)$. In this work, we denote the lower $f_0$ state as the $f_0(1500)$ resonance, since its pole mass is closer to the mass of the scalar meson $f_0(1500)$ listed in the RPP~\cite{ParticleDataGroup:2022pth}. The $f_0(1500)$ decays mostly into $2\pi$, while the $f_0(1710)$ decays mainly into the $K\bar{K}$. The $K\bar{K}$ decay branching fraction of the $f_0(1500)$ resonance is small~\cite{ParticleDataGroup:2022pth}.

In calculating the partial decay width for a scalar or tensor meson decaying into a pair of pseudoscalar mesons using the triangle diagram of  Fig.~\ref{fig:triangular}, we consider the contributions of different charges of $V_1$ and $V_2$ in the triangle loop, thus we need an extra factor $C_1$, which is a coefficient from isospin projection. It can be evaluated from the unitary normalization as following: 
   \begin{eqnarray}
      \ket{\rho \rho}_{I=0}&=&-\frac{1}{\sqrt{6}}\ket{\rho^0 \rho^0 +\rho^+ \rho^-+\rho^- \rho^+}, \\
      \ket{K^*\bar{K}^*}_{I=0}&=&-\frac{1}{\sqrt{2}}\ket{K^{*+}K^{*-}+K^{*0}\bar{K}^{*0}}, \\
      \ket{\omega \omega}_{I=0}&=&\frac{1}{\sqrt{2}}\ket{\omega \omega}, \\
      \ket{\phi \phi}_{I=0}&=&\frac{1}{\sqrt{2}}\ket{\phi \phi}, \\
      \ket{\omega \phi}_{I=0}&=&\ket{\omega \phi}, \\
      \ket{\rho K^{*}}_{I=\frac{1}{2}}&=&\frac{1}{\sqrt{3}}\ket{\rho^0 K^{*0}}-\sqrt{\frac{2}{3}}\ket{\rho^- K^{*+}}, \\
      \ket{K^* \omega}_{I=\frac{1}{2}}&=&\ket{K^{*0} \omega}, \\
      \ket{K^* \phi}_{I=\frac{1}{2}}&=&\ket{K^{*0} \phi}, \\
      \ket{K^*\bar{K}^*}_{I=1}&=&\frac{1}{\sqrt{2}}\left( \ket{K^{*0}\bar{K}^{*0}}-\ket{K^{*+}K^{*-}}\right), \\
      \ket{\rho \omega}_{I=1}&=&\ket{\rho \omega}, \\
      \ket{\rho \phi}_{I=1}&=&\ket{\rho^0 \phi}.
   \end{eqnarray}

For the $VPP$ vertex in the triangular diagram, we take the vector-pseudoscalar-pseudoscalar effective Lagrangian as used in Refs.~\cite{Molina:2008jw,Geng:2008gx,Wang:2019niy,Wang:2021jub,Wang:2022pin,Molina:2010bk}.
\begin{equation}
   \begin{aligned}
      \mathcal{L}_{VPP}=-ig \langle V_\mu \left[ P,\partial^\mu P \right]\rangle ,
   \end{aligned}
   \label{eq:VPP}
\end{equation}
\noindent
with $g = M_V/2f$. In this work, we take $M_V=M_{\rho}$ and $f=93$ MeV. The $\langle \rangle $ stands for trace in the SU(3) flavor space. $V_\mu$ and $P$ represent the vector nonet and pseudoscalar octet, respectively. They take the usual forms as
\begin{equation}
   \begin{aligned}
      V_\mu=
      \begin{pmatrix}
         \frac{\omega+\rho^0}{\sqrt{2}} & \rho^+ & K^{*+}\\
         \rho^- & \frac{\omega-\rho^0}{\sqrt{2}} & K^{*0}\\
         K^{*-} & \bar{K}^{*0} & \phi
      \end{pmatrix}_\mu,
   \end{aligned}
   \label{eq:V}
\end{equation}
and
\begin{equation}
   \begin{aligned}
      P=
      \begin{pmatrix}
         \frac{\eta}{\sqrt{6}}+\frac{\pi^0}{\sqrt{2}} & \pi^+ & K^+\\
         \pi^- & \frac{\eta}{\sqrt{6}}-\frac{\pi^0}{\sqrt{2}} & K^0\\
         K^- & \bar{K}^0 & -\sqrt{\frac{2}{3}}\eta
      \end{pmatrix}.
   \end{aligned}
   \label{eq:P}
\end{equation}

Using the interaction Lagrangian in Eq.~\eqref{eq:VPP} and the explicit forms for $V_\mu$ and $P$ in Eqs.~\eqref{eq:V} and \eqref{eq:P}, one can easily obtain $t_{V_1P_1P_t}$ and $t_{V_2P_2P_t}$ for the up and down $VPP$ vertices in Fig.~\ref{fig:triangular} as
\begin{eqnarray}
 t_{V_1P_1P_t} &=& -ig C_2\epsilon(V_1) \cdot (2k_1-q), \label{eq:vpp1} \\
 t_{V_2P_2P_t} &=& -ig C_3\epsilon(V_2) \cdot (k_2-k_1+q),   \label{eq:vpp2}
\end{eqnarray}
where $k_1$, $k_2$, and $q$ are the four-momentum of pseudoscalar mesons $P_1$ and $P_2$ and vector meson $V_1$, respectively. The coefficients $C_2$ and $C_3$ are isospin factors and they can be obtained from Eqs.~(\ref{eq:VPP})-(\ref{eq:P}).

Next, we can write down the decay amplitude for the triangular diagram shown in Fig.~\ref{fig:triangular} as
\begin{eqnarray}
t &=& Cg_{RV_1V_2} g^2  \int\frac{d^4q}{(2\pi)^4}t_a F^2 \frac{1}{q^2-M_{V_1}^2+i\epsilon}   \nonumber \\
&& \!\!\!\!  \times \frac{1}{(P-q)^2-M_{V_2}^2+i\epsilon} \frac{1}{(q-k_1)^2-m_{P_t}^2+i\epsilon}, \label{eq:tmatrix}
\end{eqnarray}
where $C$ is a global factor including the isospin factor $C_i$. The values of $C$ for all the different decay channels considered in this work are shown in the Appendix. Here, $P$ is the four-momentum of the decaying resonance $R$, and
\begin{eqnarray}
t_a &=& \mathcal{P}_J^{*}(V_1V_2) \epsilon(V_1) \cdot (2k_1-q)  \nonumber \\
&& \times \epsilon(V_2) \cdot (k_2-k_1+q).
   \label{eq:t_a}
\end{eqnarray}

To consider the off-shell effects of the exchanged pseudoscalar meson, we also consider a form factor $F$ for the $VPP$ vertices~\cite{Molina:2008jw,Geng:2008gx,Wang:2021jub,Titov:2000bn,Titov:2001yw,Wang:2022pin}
\begin{equation}
   \begin{aligned}
      F=\frac{\Lambda^2_t - m_{P_t}^2}{\Lambda^2_t-(E_1-\tilde{q}^0)^2 + |\vec{q}-\vec{k}|^2},
   \end{aligned}
   \label{eq:form}
\end{equation}
\noindent
with $\tilde{q}^0 = (M^2+M_{V_1}^2-M_{V_2}^2)/2M$, and
\begin{eqnarray}
E_1 &=& \frac{M^2+m_{P_1}^2-m_{P_2}^2}{2M}, \\
E_2 &=& \frac{M^2+m_{P_2}^2-m_{P_1}^2}{2M}, \\
|\vec{k}_1| &=& \sqrt{E_1^2-m_{P_1}^2}, \\
|\vec{k}_2| &=& \sqrt{E_2^2-m_{P_2}^2},
\end{eqnarray}
\noindent
where $M$ is the mass of the decaying scalar or tensor meson, $E_1$ ($E_2$) is the energy of pseudoscalar meson $P_1$ ($P_2$) and $|\vec{k}_1|$ ($|\vec{k}_2|$) is the module of its momentum. Besides, in the calculation, we take
\begin{equation}
   \begin{gathered}
      q=(q_0,|\vec{q}|\sin{\theta} \cos{\phi}, |\vec{q}|\sin{\theta} \sin{\phi},|\vec{q}|\cos{\theta}), \\
      k_1=(E_1, |\vec{k}_1|,0,0),~~ k_2=(E_2, -|\vec{k}_1|,0,0),
   \end{gathered}
\end{equation}
\noindent
where $q$ is the running momentum in the triangular loop.

Finally, we can obtain the partial decay width for a state dynamically generated by vector-vector coupled-channel interactions to two pseudoscalar mesons using the following formula~\cite{ParticleDataGroup:2022pth}
\begin{equation}
   \begin{aligned}
   \Gamma_{R \to P_1P_2}=\frac{1}{2J+1} \frac{|\vec{k}_1|}{8 \pi M^2} \sum_{pol.}|t|^2,
   \end{aligned}
   \label{eq:gamma}
\end{equation}
\noindent
where we sum the polarization of the intermediate vector mesons $V_1$ and $V_2$. The $J$ is the spin of the decaying particle. To obtain the $|t|^2$ we need to perform the loop integration of the $q$ in the triangular loop. We give the expressions for the calculation of $|t|^2$ in the Appendix.

\section{Numerical results and discussions}~\label{numericalresults}

We calculate the partial decay widths of the two pseudoscalar meson decay modes $\pi \pi$, $K \bar{K}$, and $\eta \eta$ of $f_2(1270)$, $f_2'(1525)$, $f_0(1500)$, and $f_0(1710)$, $K \bar{K}$ and $\pi \eta$ of $a_0(1710)$, and $K \pi$ and $K \eta$ of $K_2^*(1430)$. In our model, there is only one free parameter, the cutoff $\Lambda_t$. By adjusting its value, we can reproduce the partial decay widths of the $f_2(1270) \to  \pi \pi$ and $f_2(1270) \to  \eta \eta$ processes, for which their experimental measurements have smaller uncertainties compared to the other decays~~\cite{ParticleDataGroup:2022pth}. Hence, we take $\Lambda_t = 1350$ MeV.~\footnote{In fact, for different exchanged meson, one can take different $\Lambda_t$. To minimize the number of free parameters, we take the same value for $\Lambda_t$ for all the exchanged mesons.}

Indeed, the choice of the cutoff affects the obtained partial decay widths, especially when the exchanged meson has a large mass. We consider the uncertainty originating from $\Lambda_t$ to check our result, and we find that it does not change our final conclusion in any significant way. Moreover, to reduce the uncertainty arising from the introduction of the form factor, we also study the ratios between different partial decay widths, which are less affected by $\Lambda_t$. For doing this, we define
\begin{eqnarray}
      R_{f_2(1270)}^{K\bar{K}/\pi\pi} &=& \frac{\Gamma_{f_2(1270) \to K\bar{K}}}{\Gamma_{f_2(1270) \to \pi \pi}}, ~~
     R_{f_2(1270)}^{\eta\eta/\pi\pi} = \frac{\Gamma_{f_2(1270) \to \eta \eta}}{\Gamma_{f_2(1270) \to \pi \pi}}, \nonumber \\
    R_{f'_2(1525)}^{\pi\pi/K\bar{K}} &=& \frac{\Gamma_{f_2'(1525) \to \pi \pi}}{\Gamma_{f_2'(1525) \to K \bar{K}}}, ~~
     R_{f'_2(1525)}^{\eta \eta /K\bar{K}}= \frac{\Gamma_{f_2'(1525) \to \eta \eta}}{\Gamma_{f_2'(1525) \to K \bar{K}}}, \nonumber \\
      R_{K_2^*(1430)}^{K \eta/K \pi} &=& \frac{\Gamma_{K_2^*(1430) \to K \eta}}{\Gamma_{K_2^*(1430) \to K \pi} }, ~~
       R_{f_0(1500)}^{K\bar{K}/\pi \pi} = \frac{\Gamma_{f_0(1500) \to K\bar{K}}}{\Gamma_{f_0(1500) \to \pi \pi}}, \nonumber  \\
      R_{f_0(1500)}^{\eta \eta /\pi \pi} &=& \frac{\Gamma_{f_0(1500) \to \eta \eta}}{\Gamma_{f_0(1500) \to \pi \pi}}, ~~
      R_{f_0(1710)}^{\pi \pi/K\bar{K}}=\frac{\Gamma_{f_0(1710) \to \pi \pi}}{\Gamma_{f_0(1710) \to K \bar{K}}}, \nonumber \\
     R_{f_0(1710)}^{\eta \eta/K\bar{K}} &=& \frac{\Gamma_{f_0(1710) \to \eta \eta}}{\Gamma_{f_0(1710) \to K \bar{K}}}, ~~
      R_{a_0(1710)}^{\pi \eta/K\bar{K}}=\frac{\Gamma_{a_0(1710) \to \pi \eta}}{\Gamma_{a_0(1710) \to K\bar{K}}}. \nonumber
\end{eqnarray}

We will begin by summarizing all the numerical results for the partial decay widths and these above ratios in Tables~\ref{Tab:w} and \ref{Tab:r}, respectively. The theoretical uncertainties are determined by changing  the cutoff parameter in the range of $\Lambda_t \pm 30$ MeV. Indeed, one can see that the relative uncertainties for these ratios are smaller than the partial decay widths.

Next, we will discuss these partial decay widths and their ratios one by one.

\begin{table}[htbp]\footnotesize
 \centering
   \renewcommand\arraystretch{2.5}
   \tabcolsep=2pt
   \caption{Partial decay widths (in units of MeV) for different decay channels of $f_2(1270)$, $f_2'(1525)$, $f_0(1710)$, $K_2^*(1430)$, $f_0(1710)$, $a_0(1710)$, and $f_0(1500)$. }
   \scalebox{0.85}{
   \begin{tabular}{c|c|c|c}
       \toprule[1pt]
      \toprule[1pt]
      Decay mode & This work & RPP~\cite{ParticleDataGroup:2022pth} & Exp.\\
      \hline
       $f_2(1270) \to \pi \pi$ & $141.1^{+68.8}_{-48.0}$  & $157.2_{-1.1}^{+4.0}$  & $157.0_{-1.0}^{+6.0}$~\cite{Longacre:1986fh} \quad $152\pm 8$~\cite{Shchegelsky:2006et}\\
       $f_2(1270) \to K\bar{K}$&$15.63^{+10.29}_{-6.68}$& $8.6 \pm 0.8$ & $9.0_{-0.3}^{+0.7}$~\cite{Longacre:1986fh} \quad $7.5 \pm 2.0$~\cite{Shchegelsky:2006et}\\
       $f_2(1270) \to \eta \eta$ & $0.87^{+0.61}_{-0.39}$ & $0.75\pm0.14$ & $1.0\pm0.1$~\cite{Longacre:1986fh} \quad $1.8\pm0.4$~\cite{Shchegelsky:2006et}\\
       \hline
        $f_2'(1525) \to \pi \pi$ & $5.5^{+1.0}_{-1.0}$& $0.71\pm0.14$ &$1.4_{-0.7}^{+1.0}$~\cite{Longacre:1986fh} \quad $0.2_{-0.2}^{+1.0}$~\cite{Shchegelsky:2006et}\\
         $f_2'(1525) \to  K \bar{K}$ & $17.6^{+10.1}_{-6.9}$&$75\pm4$ &$63_{-5}^{+6}$~\cite{Longacre:1986fh}\\
         $f_2'(1525) \to \eta \eta $ & $6.6^{+4.3}_{-2.9}$&$9.9\pm1.9$ &$24_{-1}^{+3}$~\cite{Longacre:1986fh}\quad $5.0\pm8$~\cite{Shchegelsky:2006et}  \\
        \hline
          $K_2^*(1430) \to K\pi$ &$34.6^{+16.5}_{-11.7}$ & $53.2 \pm 2.9$ \\
          $K_2^*(1430) \to K\eta$ &$2.62^{+1.65}_{-1.09}$ & $0.16^{+0.36}_{-0.11}$\\
          \hline
         $f_0(1710) \to \pi \pi$ & $5.76^{+2.04}_{-1.31}$& $5.85^{+0.30}_{-4}$ \\
         $f_0(1710) \to  K \bar{K}$ & $57.9^{+28.8}_{-20.5}$& $54^{+18}_{-19}$\\
          $f_0(1710) \to \eta \eta $ & $32.7^{+18.8}_{-12.9}$& $33\pm 18$\\
          \hline
          $a_0(1710) \to K\bar{K}$ & $40.2^{+26.8}_{-17.6}$ & & \\
          $a_0(1710) \to K \eta$ & $38.4^{+19.9}_{-14.2}$ & &\\
          \hline
          $f_0(1500) \to \pi \pi$ & $632^{+269}_{-198}$& $37\pm12$&\\
          $f_0(1500) \to K\bar{K}$ & $51.0^{+30.6}_{-20.5}$ &$9.2\pm3.0$ &\\
          $f_0(1500) \to \eta \eta$ & $0.54^{+0.37}_{-0.24}$ &$6.5\pm2.2$ &\\
      \bottomrule[1pt]
      \bottomrule[1pt]
   \end{tabular}
   }
   \label{Tab:w}
\end{table}

\begin{table}[htbp]\footnotesize
    \centering
   \renewcommand\arraystretch{3}
   \tabcolsep=2pt
   \caption{Branching ratios for different decay channels of $f_2(1270)$, $f_2'(1525)$, $f_0(1710)$, $K_2^*(1430)$, $a_0(1710)$, and $f_0(1500)$.}
   \scalebox{0.85}{
   \begin{tabular}{c|c|c|c|c}
    \toprule[1pt]
    \toprule[1pt]
    \multicolumn{2}{c|}{Ratio} & This work  & RPP~\cite{ParticleDataGroup:2022pth} & Exp.\\
    \hline
    \multirow{2}{*}{$f_2(1270)$} & $R_{f_2(1270)}^{K\bar{K}/\pi\pi}$ & $0.11 \pm 0.02$ &  $0.0041_{-0.005}^{+0.004}$ &$0.02-0.06$~\cite{OBELIX:2002lhi,Etkin:1981sg,CERN-CRACOW-MUNICH:1981atg,CERN-CollegedeFrance-Madrid-Stockholm:1979etn,BES:2004zql,BARI-BONN-CERN-GLASGOW-LIVERPOOL-MILAN-VIENNA:1980wqx,Martin:1979gm,Polychronakos:1978ur,Emms:1975kt,Aderholz:1969ya}\\
    \cline{2-5}
    & $R_{f_2(1270)}^{\eta \eta/\pi \pi}$ & $0.006 \pm 0.001$ &  & $0.003\pm0.001$~\cite{WA102:2000lao}\\
    \hline
    \multirow{2}{*}{$f_2'(1525)$} &  $R_{f_2'(1525)}^{\pi \pi/K\bar{K}}$& $0.31^{+0.11}_{-0.08}$  & $0.0094\pm0.0018$ &$0.075\pm0.035$~\cite{DM2:1987vpm}\\
    \cline{2-5}
    & $R_{f_2'(1525)}^{\eta \eta/K\bar{K}}$ &$0.37^{+0.03}_{-0.02}$ &  $0.115\pm0.028$ &\thead{ $0.119\pm0.015\pm0.036$~\cite{Binon:2007zz} \\ $0.11\pm0.04$~\cite{Prokoshkin:1990eu}}\\
    \hline
    \multirow{2}{*}{$f_0(1710)$} &$R_{f_0(1710)}^{\pi \pi/K\bar{K}}$& $0.10 \pm 0.02$&$0.23\pm0.05$ &\thead{$0.64\pm0.27\pm0.18$~\cite{BaBar:2018uqa}\\ $0.41_{-0.17}^{+0.11}$~\cite{Ablikim:2006db}\\$0.2\pm0.024\pm0.036$~\cite{WA102:1999fqy}\\ $0.39\pm0.14$~\cite{WA76:1991kef}\\$0.32\pm0.14$~\cite{Albaladejo:2008qa} \\ $5.8_{-5.5}^{+9.1}$~\cite{Anisovich:2001ay}}\\
    \cline{2-5}
    &$ R_{f_0(1710)}^{\eta \eta/K\bar{K}}$& $0.56 \pm 0.04$  &&\thead{$0.48\pm0.15$~\cite{WA102:2000lao}\\ $0.46_{-0.38}^{+0.70}$~\cite{Anisovich:2001ay}}\\
    \hline
    $K_2^*(1430)$ &$R_{K_2^*(1430)}^{K\eta/K\pi}$ &$0.075 \pm 0.009$  &$0.0030^{+0.0070}_{-0.0020}$ & $ 0 \pm 0.0056$~\cite{Aston:1987ey}\\
    \hline
    $a_0(1710)$ & $R_{a_0(1710)}^{K\eta/K\bar{K}}$&$0.96^{+0.14}_{-0.09}$& & \\
    \hline
    \multirow{2}{*}{\thead{\\ \\ \\ \\$f_0(1500)$}} &$R_{f_0(1500)}^{K\bar{K}/\pi \pi}$&$0.08\pm 0.01$&$0.236\pm0.026$&\thead{$0.25\pm0.03$~\cite{OBELIX:2002lhi} \\
    $0.33\pm0.03\pm0.07$~\cite{WA102:1999fqy}\\$0.16\pm 0.05$~\cite{Anisovich:2001ay}\\ $0.19\pm0.07$~\cite{Abele:1998qd}\\$0.20\pm0.08$~\cite{CrystalBarrel:1996sqr}}\\
     \cline{2-5}
    &$R_{f_0(1500)}^{\eta \eta/\pi \pi}$&($8.6\pm1.5)\times 10^{-4}$&$0.175\pm0.027$&\thead{$0.18\pm0.03$~\cite{WA102:2000lao} \\ $0.11\pm0.03$~\cite{Anisovich:2001ay}\\ $0.157\pm0.060$~\cite{CrystalBarrel:1995dzq}\\$0.080\pm0.033$~\cite{CrystalBarrel:2002qpf}\\ $0.078\pm0.013$~\cite{Abele:1996si} \\ $0.230\pm0.097$~\cite{CrystalBarrel:1995fiy}}\\
    \bottomrule[1pt]
    \bottomrule[1pt]  
   \end{tabular}
   }
   \label{Tab:r}
\end{table}

\subsection{Tensor mesons: $f_2(1270)$, $f'_2(1525)$, and $K^*_2(1430)$}

Table~\ref{Tab:w} shows that our numerical results for $f_2(1270)$ agree well with the experimental results quoted in the RPP~\cite{ParticleDataGroup:2022pth} within uncertainties, which partly support the $\rho \rho$ molecular nature of the $f_2(1270)$ resonance. The predominant decay mode is $f_2(1270) \to \pi\pi$, while the $f_2(1270) \to \eta \eta$ decay width is very small.

For the decay of $f_2'(1525) \to \eta \eta$, our result is consistent with the experimental measurements~\cite{ParticleDataGroup:2022pth,Shchegelsky:2006et}, but for the processes $f_2'(1525) \to \pi\pi $ and $f_2'(1525) \to K\bar{K}$, our results are different from the experimental ones~\cite{Longacre:1986fh,Shchegelsky:2006et}. The obtained $\Gamma_{f'_2(1525) \to \pi\pi}$ is larger than the experimental data, while the obtained $\Gamma_{f'_2(1525) \to K \bar{K}}$ is smaller than the experimental measurement. Obviously, further investigations both on theoretical and experimental sides are needed to deepen our understanding of the nature of $f_2'(1525)$.

The  experimental information about $K_2^*(1430)$ is very limited~\cite{ParticleDataGroup:2022pth}. Most of its properties were extracted from the $KN$ scattering experiments in 1970s and 1980s. With the total width~\footnote{The total width of $\Gamma_{K^*_2(1430)}$ is calculated by averaging the widths of charged and neutral $K^*_2(1430)$.} $\Gamma_{K^*_2(1430)} = 104.5 \pm 5.5$ MeV and the branching fraction $Br[K^*_2(1430) \to K \pi] = (49.9 \pm 1.2)\%$ as quoted in the RPP~\cite{ParticleDataGroup:2022pth}, one can obtain the partial decay width of $K^*_2(1430) \to K \pi$
\begin{equation}
   \Gamma_{K_2^*(1430) \to K\pi} = 52.1 \pm 3.0 ~\rm{MeV},
   \label{eq:w-kpi}
\end{equation}
which is consistent with our theoretical results $34.6^{+16.5}_{-11.7}$ MeV within uncertainties. Similarly, according to the branching fraction $Br[K^*_2(1430) \to K \eta] = (0.15^{+0.34}_{-0.10})\%$~\cite{ParticleDataGroup:2022pth}, we obtain
\begin{equation}
   \Gamma_{K_2^*(1430) \to K\eta} = 0.16_{-0.10}^{+0.36}~\rm{MeV},
   \label{eq:w-keta}
\end{equation}
which is smaller than our result as shown in Table~\ref{Tab:w}.

The theoretical $\Gamma_{K_2^*(1430) \to K\pi}$ and the ratio $R_{K_2^*(1430)}^{K \eta/K \pi}$ are consistent with the previous theoretical calculations within uncertainties~\cite{Geng:2009gb}. However, the obtained $\Gamma_{K_2^*(1430) \to K \eta}$ is much different from the experimental result~\cite{ParticleDataGroup:2022pth}.

\subsection{Scalar mesons: $f_0(1710)$, $f_0(1500)$, and $a_0(1710)$}

For the scalar meson $f_0(1710)$, the corresponding experimental measurements of its decays are very few~\cite{ParticleDataGroup:2022pth}. Our theoretical results are in agreement with the previous theoretical analyses~\cite{Longacre:1986fh,Albaladejo:2008qa}. With $Br[f_0(1710) \to \pi \pi] = 0.039^{+0.002}_{-0.024}$~\cite{Longacre:1986fh}, $Br[f_0(1710) \to K \bar{K}] = 0.36\pm0.12$~\cite{Albaladejo:2008qa}, $Br[f_0(1710) \to \eta \eta] = 0.22\pm0.12$~\cite{Albaladejo:2008qa}, and $\Gamma_{f_0(1710)} = 150_{-10}^{+12}$ MeV~\cite{ParticleDataGroup:2022pth}, one can obtain
\begin{eqnarray}
 \Gamma_{f_0(1710) \to \pi \pi} &=& 5.9^{+0.3}_{-4.0}~\rm{MeV},\\
   \Gamma_{f_0(1710) \to K\bar{K}} &=& 54_{-19}^{+18}~\rm{MeV},\\
   \Gamma_{f_0(1710) \to \eta \eta} &=& 33 \pm18 ~\rm{MeV} ,
\end{eqnarray}
which are consistent with our numerical results shown in Table~\ref{Tab:w}. Meanwhile, on can also obtain
\begin{eqnarray}
  R_{f_0(1710)}^{\eta \eta/K\bar{K}} &=& 0.56 \pm 0.04,
\end{eqnarray}
which is in agreement with the experimental measurements~\cite{WA102:2000lao,Anisovich:2001ay}. This value is slightly larger than the one, $0.294 \pm 0.048$, obtained in Ref.~\cite{Wang:2021jub}.

For the isovector $a_0(1710)$, there are no experimental data on its decays. We show the partial decay widths of $a_0(1710)$ in Table~\ref{Tab:w}. It is expected that these results can be tested by future experimental measurements, for instance, in the decays of charmonium states~\cite{Ding:2023eps,Ding:2024lqk}.

For the $f_0(1500)$, the obtained partial decay widths are shown in Table~\ref{Tab:w}, which are much larger than the experimental data~\footnote{There are many difficulties in studying the $f_0(1500)$, because of the strong interference among $f_0(500)$, $f_0(1370)/f_0(1500)$, and $f_0(1710)$.} quoted in the RPP~\cite{ParticleDataGroup:2022pth}, especially for its decay to the $\pi \pi$ channel. This is the reason that, in the original work of Ref.~\cite{Geng:2008gx}, the lower dynamically generated $f_0$ state was associated to the $f_0(1370)$ resonance rather than the $f_0(1500)$ resonance. However, because of the limited experimental information for $f_0(1370)$ and $f_0(1500)$~\cite{ParticleDataGroup:2022pth}, it is very difficult to make a definite conclusion at this time.

The branching decay ratios of $f_0(1500)$ are also obtained as shown in Table~\ref{Tab:r}. In addition, the relevant experimental measurements for the $f_0(1370)$ are listed in the following:

\begin{equation}
   R^{K\bar{K}/\pi\pi}_{f_0(1370)} = 
   \begin{array}{l}
   0.91 \pm 0.20 \; \text{\cite{OBELIX:2002lhi}} \\
   0.46 \pm 0.15 \pm 0.11 \; \text{\cite{WA102:1999fqy}} \\
   0.12 \pm 0.06 \; \text{\cite{Anisovich:2001ay}} \\
   0.08 \pm 0.08 \; \text{\cite{BES:2004twe}}
   \end{array}
\end{equation}

One can see that these experimental results are  not consistent with each other. The $R^{K\bar{K}/\pi\pi}_{f_0(1370)}$ indicates that its partial decay width to the $\pi \pi$ channel is lager than the one to the $K\bar{K}$ channel. However, the fractions of $Br[f_0(1370) \to \pi \pi] = 0.26 \pm 0.06$ and $Br[f_0(1370) \to K\bar{K}] = 0.35 \pm 0.13$ were obtained in Ref.~\cite{Bugg:1996ki}. This indicates the $K\bar{K}$ channel is also important for the $f_0(1370)$ resonance. Besides, smaller upper limits for the fraction of $Br[f_0(1370) \to \pi \pi]$ were measured in Refs.~\cite{Ochs:2013gi,CrystalBarrel:1994doj,Gaspero:1992gu}.

We note that it is not yet very clear to which state to associate the lower dynamically generated  scalar state in the $\rho\rho$ channel, $f_0(1370)$ or $f_0(1500)$, given the large uncertainties on the masses and widths of $f_0(1370)$ and the fact that the predicted partial decay widths do not agree with the experimental data in either assignment, although in this work we tentatively refer to it as $f_0(1500)$. More accurate experimental data and/or careful analyses of the relevant experimental data are strongly encouraged.

\section{Summary}~\label{Summary}

We investigated the two-body strong decays of the $f_0(1500)$, $f_0(1710)$, $a_0(1710)$, $f_2(1270)$, $f_2'(1525)$, and $K_2^*(1430)$ resonances, which are dynamically generated from the $s$-wave interactions between two vector mesons. The contributions from triangular diagrams are analyzed, and the interference effects among different vector-vector channels are accounted for. Our theoretical predictions agree with the experimental data within uncertainties, e.g.: $f_2(1270) \to \pi \pi$, $K\bar{K}$, $\eta \eta$, $f_2'(1525) \to \eta \eta$, and $f_0(1710) \to K\bar{K}$, $\eta \eta$. Based on our calculations, the $f_2(1270)$ is identified as a $\rho \rho$ bound state. However, some theoretical predictions deviate from the experimental data, indicating the need for further studies on both theoretical and experimental sides to resolve these discrepancies, particularly those of $f_0(1500)$. Future research will contribute to a deeper understanding of the nature of these tensor and scalar resonances.

\section*{acknowledgments}

We would like to thank Prof. Eulogio Oset, Prof. Xiang Liu, and Prof. En Wang for useful discussions and careful reading the manuscript. This work is partly supported by the National Key R\&D Program of China under Grant No. 2023YFA1606703, and by the National Natural Science Foundation of China under Grant Nos. 12075288, 12435007 and 12361141819. It is also supported by the Youth Innovation Promotion Association CAS.

\section*{Appendix: Coefficients $C$}

The obtained coefficients $C$ where the isospin factor $C_i$ are taken into account are presented in the following Tables~\ref{Tab:C-f}-\ref{Tab:C-a0}. Note that there is an extra factor $1/2$ in Eq.(~\ref{eq:gamma}) for the cases of $\pi^0 \pi^0$ and $\eta \eta$ in the final states.

\begin{table}[htbp]\footnotesize
   \centering
   \renewcommand\arraystretch{2}
   \tabcolsep=1.5pt
   \caption{Coefficients $C$ for $f_2(1270)$, $f_2'(1525)$, $f_0(1710)$, and $f_0(1500)$, where only $K^*\bar{K}^*$ and $\rho \rho$ is considered for 
   $f_2(1270)$ and $f_0(1500)$ because their couplings to other channels are small compared with those of $\rho \rho$ and $K^*\bar{K}^*$ channels.}
   \begin{tabular}{cccccccccccc}
      \toprule[1pt]
      \toprule[1pt]
       $V_1$ & $V_2$ & $P_t$ &  $P_1$ & $P_2$ &$C$ & $V_1$ & $V_2$ & $P_t$ &  $P_1$ & $P_2$ &$C$\\
       \hline
       \multicolumn{12}{c}{$\pi \pi$}\\
       \hline
       $K^{*+}$ & $K^{*-}$ & $K^+$ & $\pi^0$ & $\pi ^0$ & $-\frac{1}{\sqrt{2}}$ & $K^{*+}$ & $K^{*-}$ & $K^0$ & $\pi^+$ & $\pi^-$ & $-\frac{1}{\sqrt{2}}$\\
       \hline
       $K^{*0}$ & $\bar{K}^{*0}$ & $K^0$ & $\pi^0$ & $\pi^0$ & $-\frac{1}{\sqrt{2}}$ & $\bar{K}^{*0}$ & $K^{*0}$ & $K^-$ & $\pi^+$ & $\pi^-$ & $-\frac{1}{\sqrt{2}}$\\
       \hline
       $\rho^+$ & $\rho^-$ & $\pi^+$ & $\pi^0$ & $\pi^0$ & $-4\sqrt{\frac{2}{3}}$ & $\rho^0$ & $\rho^0$ & $\pi^-$ & $\pi^+$ & $\pi^-$ & $-2\sqrt{\frac{2}{3}}$\\
       \hline
        & & & & & & $\rho^+$ & $\rho^-$ & $\pi^0$ & $\pi^+$ & $\pi^-$ & $-2\sqrt{\frac{2}{3}}$\\
        \hline
        \multicolumn{12}{c}{$K\bar{K}$}\\
        \hline
        $K^{*+}$ & $K^{*-}$ & $\pi^0$ & $K^+$ & $K^-$ & $-\frac{1}{2\sqrt{2}}$ & $K^{*+}$ & $K^{*-}$ & $\pi^+$ & $K^0$ & $\bar{K}^0$ & $-\frac{1}{\sqrt{2}}$\\
        \hline
        $K^{*0}$ & $\bar{K}^{*0}$ & $\pi^-$ & $K^+$ & $K^-$ & $-\frac{1}{\sqrt{2}}$ & $K^{*0}$ & $\bar{K}^0$ & $\pi^0$ & $K^0$ & $\bar{K}^0$ & $-\frac{1}{2\sqrt{2}}$\\
        \hline
        $K^{*+}$ & $K^{*-}$ & $\eta$ & $K^+$ & $K^-$ & $-\frac{3}{2\sqrt{2}}$ & $K^{*0}$ & $\bar{K}^{*0}$ & $\eta$ & $K^0$ & $\bar{K}^0$ & $-\frac{3}{2\sqrt{2}}$\\
        \hline
        $\rho^0$ & $\rho^0$ & $K^-$ & $K^+$ &  $K^-$ & $-\sqrt{\frac{1}{6}}$ & $\rho^0$ & $\rho^0$ & $\bar{K}^0$ & $K^0$ & $\bar{K}^0$ & $-\sqrt{\frac{1}{6}}$\\
        \hline7
        $\rho^+$ & $\rho^-$ & $\bar{K}^0$ & $K^+$ & $K^-$ & $-\sqrt{\frac{2}{3}}$ & $\rho^+$ & $\rho^-$ & $K^+$ & $\bar{K}^0$ & $K^0$ & $-\sqrt{\frac{2}{3}}$ \\
        \hline
        $\omega$ & $\omega$ & $K^-$ & $K^+$ & $K^-$ & $\frac{1}{\sqrt{2}}$ & $\omega$ & $\omega$ & $\bar{K}^0$ & $K^0$ & $\bar{K}^0$ & $\frac{1}{\sqrt{2}}$\\
        \hline
        $\omega$ & $\phi$ & $K^-$ & $K^+$ & $K^-$ & $-\frac{1}{\sqrt{2}}$ & $\omega$ & $\phi$ & $\bar{K}^0$ & $K^0$ & $\bar{K}^0$ & $-\frac{1}{\sqrt{2}}$\\
        \hline
        $\omega$ & $\phi$ & $K^+$ & $K^-$ & $K^+$ & $-\frac{1}{\sqrt{2}}$ & $\omega$ & $\phi$ & $K^0$ & $\bar{K}^0$ & $K^0$ & $-\frac{1}{\sqrt{2}}$\\
        \hline
        $\phi$ & $\phi$ & $K^-$ & $K^+$ & $K^-$ & $\sqrt{2}$ & $\phi$ & $ \phi$ & $K^0$ & $\bar{K}^0$ & $K^0$ & $\sqrt{2}$\\
        \hline
        \multicolumn{12}{c}{$\eta \eta$}\\
        \hline
        $K^{*+}$ & $K^{*-}$ & $K^+$ & $\eta$ & $\eta$ & $-\frac{3}{\sqrt{2}}$ & $K^{*0}$ & $K^{*-}$ & $K^+$ & $\eta$ & $\eta$ & $-\frac{3}{\sqrt{2}}$\\
      \bottomrule[1pt]
      \bottomrule[1pt]
   \end{tabular}
   \label{Tab:C-f}
\end{table}

\begin{table}[htbp]\footnotesize
   \centering
   \renewcommand\arraystretch{2}
   \tabcolsep=1.5pt
   \caption{Coefficients $C$ for $K_2^*(1430)$.}
   \begin{tabular}{cccccccccccc}
      \toprule[1pt]
      \toprule[1pt]
       $V_1$ & $V_2$ & $P_t$ &  $P_1$ & $P_2$ &$C$ & $V_1$ & $V_2$ & $P_t$ &  $P_1$ & $P_2$ &$C$\\
      \hline
      \multicolumn{12}{c}{$K\pi$}\\
      \hline
      $K^{*0}$ & $\rho^0$ & $K^0$ & $\pi^0$ & $K^0$ & $-\frac{1}{2\sqrt{3}}$ & $K^{*0}$ & $\rho^0$ & $\pi^-$ & $K^+$ & $K^-$ & $-\sqrt{\frac{2}{3}}$\\
      \hline
      $K^{*+}$ & $\rho^-$ & $K^+$ & $\pi^0$ & $K^0$ & $\frac{1}{\sqrt{3}}$ & $K^{*0}$ & $\rho^0$ & $K^+$ & $\pi^-$ & $K^+$ & $-\frac{1}{\sqrt{6}}$\\
      \hline
      $K^+$ & $\rho^-$ & $\pi^+$ & $K^0$ & $\pi^0$ & $\frac{2}{\sqrt{3}}$ & $K^{*+}$ & $\rho^-$ & $\pi^0$ & $K^+$ & $\pi^-$ & $-\sqrt{\frac{2}{3}}$\\
      \hline
      $K^{*0}$ & $\omega$ & $K^0$ & $\pi^0$ & $K^0$ & $\frac{1}{2}$ & $K^{*0}$ & $\omega$ & $K^+$ & $\pi^-$ & $K^+$ & $-\frac{1}{\sqrt{2}}$\\
      \hline
      $K^{*0}$ & $\phi$ & $K^0$ & $\pi^0$ & $K^0$ & $-\frac{1}{\sqrt{2}}$ & $K^{*0}$ & $\phi$ & $K^+$ & $\pi^-$ & $K^+$ & $1$\\
      \hline
      \multicolumn{12}{c}{$K\eta$}\\
      \hline
      $K^{*0}$ & $\rho^0$ & $K^0$ & $\eta$ & $K^0$ & $\frac{1}{2}$ & $K^{*+}$ & $\rho^-$ & $K^+$ & $\eta$ & $K^0$ & $1$\\
      \hline
      $K^{*0}$ & $\omega$ & $K^0$ & $\eta$ & $K^0$ & $-\frac{\sqrt{3}}{2}$ & $K^{*0}$ & $\phi$ & $K^0$ & $\eta$ & $K^0$ & $\sqrt{\frac{3}{2}}$\\
      \bottomrule[1pt]
      \bottomrule[1pt]
   \end{tabular}
   \label{Tab:C-K2}
\end{table}

\begin{table}[htbp]\footnotesize
   \centering
   \renewcommand\arraystretch{2}
   \tabcolsep=1.5pt
   \caption{Coefficients $C$ for $a_0(1710)$.}
   \begin{tabular}{cccccccccccc}
      \toprule[1pt]
      \toprule[1pt]
       $V_1$ & $V_2$ & $P_t$ &  $P_1$ & $P_2$ &$C$ & $V_1$ & $V_2$ & $P_t$ &  $P_1$ & $P_2$ &$C$\\
      \hline
      \multicolumn{12}{c}{$K\bar{K}$}\\
      \hline
      $K^{*0}$ & $\bar{K}^{*0}$ & $\pi^0$ & $K^0$ & $\bar{K}^0$ & $\frac{1}{2\sqrt{2}}$ & $K^{*0}$ & $\bar{K}^{*0}$ & $\pi^-$ & $K^+$ & $K^-$ & $\frac{1}{\sqrt{2}}$\\
      \hline
      $K^{*0}$ & $\bar{K}^{*0}$ & $\eta$ & $K^0$ & $\bar{K}^0$ & $\frac{3}{2\sqrt{2}}$ & $K^{*+}$ & $K^{*-}$ & $\eta$ & $K^+$ & $K^-$ & $-\frac{3}{2\sqrt{2}}$\\
      \hline
      $K^{*+}$ & $K^{*-}$ & $\pi^+$ & $K^0$ & $\bar{K}^0$ & $-\frac{1}{\sqrt{2}}$ & $K^{*+}$ & $K^{*-}$ & $\pi^0$ & $K^+$ & $K^-$ & $-\frac{1}{2\sqrt{2}}$\\
      \hline
      $\rho^0$ & $\omega$ & $\bar{K}^0$ & $K^0$ & $\bar{K}^0$ & $-\frac{1}{2}$ & $\rho^0$ & $\omega$ & $K^-$ & $K^+$ & $K^-$ & $\frac{1}{2}$\\
      \hline
      $\rho^0$ & $\omega$ & $K^0$ & $\bar{K}^0$ & $K^0$ & $-\frac{1}{2}$ & $\rho^0$ & $\omega$ & $K^+$ & $K^-$ & $K^+$ & $\frac{1}{2}$\\
      \hline
      $\rho^0$ & $\phi$ & $K^0$ & $\bar{K}^0$ & $K^0$ & $\frac{1}{\sqrt{2}}$ & $\rho^0$ & $\phi$ & $K^-$ & $K^+$ & $K^-$ & $-\frac{1}{\sqrt{2}}$\\
      \hline
      $\rho^0$ & $\phi$ & $\bar{K}^0$ & $K^0$ & $\bar{K}^0$ & $\frac{1}{\sqrt{2}}$ & $\rho^0$ & $\phi$ & $K^+$ & $K^-$ & $K^+$ & $-\frac{1}{\sqrt{2}}$\\
      \hline
      \multicolumn{12}{c}{$\pi \eta$}\\
      \hline
      $K^{*0}$ & $\bar{K}^{*0}$ & $K^0$ & $\pi^0$ & $\eta$ & $-\frac{1}{2}\sqrt{\frac{3}{2}}$ & $K^{*0}$ & $\bar{K}^{*0}$ & $K^0$ & $\eta$ & $\pi^0$ & $-\frac{1}{2}\sqrt{\frac{3}{3}}$\\
      \hline
      $K^{*+}$ & $K^{*-}$ & $K^+$ & $\pi^0$ & $\eta$ & $-\frac{1}{2}\sqrt{\frac{3}{2}}$ & $K^{*+}$ & $K^{*-}$ & $K^+$ & $\eta$ & $\pi^0$ & $-\frac{1}{2}\sqrt{\frac{3}{2}}$\\
      \bottomrule[1pt]
      \bottomrule[1pt]
   \end{tabular}
   \label{Tab:C-a0}
\end{table}

\section*{Appendix: Calculation of the triangular loop}

\begin{figure}[htbp]
   \centering
   \includegraphics[scale=0.5]{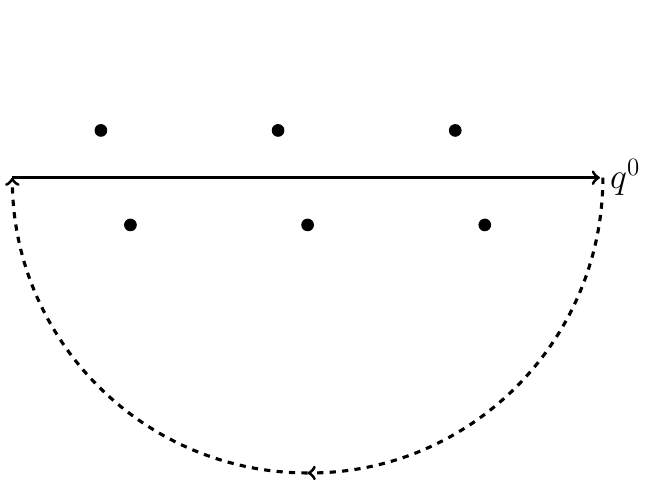}
   \caption{Contour for the triangular diagram.}
   \label{fig:path}
\end{figure}

To calculate the triangle loop integral, we first integrate the $q_0$ variable by the residue theorem as shown in Fig.~\ref{fig:path}.

Note that $F$ and $\mathcal{P}^S$ are independent of $q_0$, and only $t_a$ is a function of it, where $C$ is the global coefficient and is shown in Table~\ref{Tab:C-K2}, \ref{Tab:C-f}, and~\ref{Tab:C-a0}. Then, one can obtain
\begin{equation}
   \begin{aligned}
      t=-\frac{iCg_{RV_1V_2}g^2}{(2\pi)^3}\int^{\Lambda_t}_0 d^3\vec{q}  \frac{N}{D}P^{S*}(V_1V_2)F^2 ,
   \end{aligned}
\end{equation}
\noindent
where 
\begin{eqnarray}
      D &=& 8\omega_{V_1} \omega_{V_2}\omega_{P_t}(P_0+\omega_{V_1}+\omega_{V_2})(P_0-\omega_{V_1}+\omega_{V_2}+i\epsilon) \nonumber \\
      & &\times (P_0-\omega_{V_1}-\omega_{V_2}+i\epsilon)(E_1+\omega_{V_1}+\omega_{P_t})\nonumber \\
      & &\times (E_1-\omega_{V_1}+\omega_{P_t}+i\epsilon)(E_1-\omega_{V_1}-\omega_{P_t}+i\epsilon) \nonumber \\
      & &\times (E_2+\omega_{V_2}+\omega_{P_t})(E_2+\omega_{V_2}-\omega_{P_t}+i\epsilon) \nonumber \\
      & &\times (E_2-\omega_{V_2}-\omega_{P_t}+i\epsilon),
\end{eqnarray}
and
\begin{equation}
      N=N_1+N_2+N_3,
\end{equation}
with
\begin{eqnarray}
N_1 &=& 4\omega_{V_2}\omega_{P_t}(P_0+\omega_{V_1}+\omega_{V_2})(E_1+\omega_{V_1}+\omega_{P_t}) \times \nonumber\\
      &&(E_2+\omega_{V_2}+\omega_{P_t})(E_2+\omega_{V_2}-\omega_{P_t})(E_2-\omega_{V2}-\omega_{P_t}) \nonumber \\
      && \times t_a|_{\omega_{V_1}} \nonumber \\
      N_2 &=& 4\omega_{V_1} \omega_{P_t}(P_0-\omega_{V_1}+\omega_{V_2})(E_1+\omega_{V_1}+\omega_{P_t}) \times \nonumber \\
      &&(E_1-\omega_{V_1}+\omega_{P_t})(E_1-\omega_{V_1}-\omega_{P_t})(E_2-\omega_{V_2}-\omega_{P_t}) \nonumber \\
      && \times t_a|_{P_0+\omega_{V_2}} \nonumber \\
      N_3 &=& 4\omega_{V_1} \omega_{V_2}(P_0+\omega_{V_1}+\omega_{V_2})(P_0-\omega_{V_1}+\omega_{V_2}) \times \nonumber \\
      &&(P_0-\omega_{V_1}-\omega_{V_2})(E_1-\omega_{V_1}-\omega_{P_t})(E_2+\omega_{V_2}+\omega_{P_t}) \nonumber \\
      && \times t_a|_{E_1+\omega_{P_t}} \nonumber
\end{eqnarray}
\noindent
where $P_0=M$, $t_a|_x$ is the value of $t_a$ obtained at $q_0=x$ with Eq.(~\ref{eq:t_a}), and
\begin{equation}
   \begin{aligned}
      \omega_{V_1}&=\sqrt{|\vec{q}|^2+M_{V_1}^2},\\
      \omega_{V_2}&=\sqrt{|\vec{q}|^2+M_{V_2}^2},\\
      \omega_{P_t}&=\sqrt{|\vec{q}-\vec{k_1}|^2+m_{P_t}^2}.
   \end{aligned}
\end{equation}

On the other hand, as in Refs.~\cite{Molina:2008jw,Geng:2008gx}, to avoid the appearance of double poles and approximately account for the dispersion of vector meson masses in the convolution, we replace $\omega_{V_1}$ and $\omega_{V_2}$ in the denominator $D$ by
\begin{equation}
   \begin{aligned}
      w_{V_1} \to w_{V_1}+\frac{\Gamma_{V_1}}{4}, ~~~ w_{V_2} \to \omega_{V_2}-\frac{\Gamma_{V_2}}{4}.
   \end{aligned}
\end{equation}

\normalem
\bibliographystyle{apsrev4-1.bst}
\bibliography{reference.bib}
\end{document}